\documentclass[final,5p,times,twocolumn]{elsarticle}

\usepackage{graphicx}

\usepackage{amssymb}
\usepackage{amsmath}

\journal{Radiation Measurements}

\begin{document}

\begin{frontmatter}

\title{Search for strange quark matter and Q-balls with the SLIM experiment}

\author{Z.~SAHNOUN$^{\textrm{a,b}}$, for the SLIM collaboration\fnref{coll}}

\address{$^{a}$INFN Sez. Bologna, 40127 Bologna, Italy.}
\address{$^{b}$Astrophysics Department, C.R.A.A.G., B.P.~63 Bouzareah, 16340 Algiers, Algeria.}

\author{Talk given at the 24$^{th}$ International Conference on Nuclear Tracks in Solids,\\ Bologna, Italy, 1-5 September 2008.}

\fntext[coll]{\textbf{The SLIM Collaboration}: S. Cecchini, M. Cozzi, D. Di Ferdinando, M. Errico, F. Fabbri, G. Giacomelli, R. Giacomelli, M. Giorgini, A. Kumar, J. McDonald, G. Mandrioli, S. Manzoor, A. Margiotta, E. Medinaceli, L. Patrizii, J. Pinfold, V. Popa, I.E. Qureshi, O. Saavedra, Z. Sahnoun, G. Sirri, M. Spurio, V. Togo, C. Valieri, A. Velarde, A. Zanini.}

\begin{abstract}
We report on the search for Strange Quark Matter (SQM) and charged Q-balls with the SLIM experiment at the Chacaltaya High Altitude Laboratory (5230 m a.s.l.) from 2001 to 2005. The SLIM experiment was a $427$ m$^{2}$ array of Nuclear Track Detectors (NTDs) arranged in modules of $24 \times 24$ cm$^{2}$ area. SLIM NTDs were exposed to the cosmic radiation for 4.22 years after which they were brought back to the Bologna Laboratory where they were etched and analyzed. We estimate the properties and energy losses in matter of nuclearites (large SQM nuggets), strangelets (small charged SQM nuggets) and Q-balls; and discuss their detection with the SLIM experiment. The flux upper limits in the CR of such downgoing particles are at the level of $1.3\cdot10^{-15}$/cm$^{2}$/s/sr (90\% CL).
\end{abstract}


\end{frontmatter}

\section{Introduction/Scope}
The SLIM (Search for LIght magnetic Monopoles) experiment was a large array ($427$ m$^{2}$) of Nuclear Track Detectors (NTDs) operated at the high altitude Chacaltaya laboratory in Bolivia (5230 m a.s.l.) from 2001 to 2005. The experiment was designed to search for light and intermediate mass magnetic monopoles (Cecchini et al., 2001a) predicted by some grand unified theories (GUT) and supersymmetric models and could have been produced in the Early Universe. It was also sensitive to Strange Quark Matter (SQM) and Q-balls (Balestra et al., 2006 , Cecchini et al., 2005).\\ 
Composed of approximately the same number of up, down and strange quarks SQM was conjectured to be the ground state of nuclear matter (Witten, E. , 1984) and may be stable for all baryon numbers ranging from ordinary nuclei to neutron stars ($A$$\sim$$10^{57}$). Large lumps of strange quark matter hypothetically present in the cosmic radiation are generally called Nuclearites. They are believed to be neutralized by an electron cloud forming a sort of atom and to interact with the surrounding medium via elastic and quasi elastic collisions (De Rujula and Glashow, 1984, De Rujula, 1985).

\begin{figure}
\centering
\resizebox{7.1cm}{6.7cm}{\includegraphics{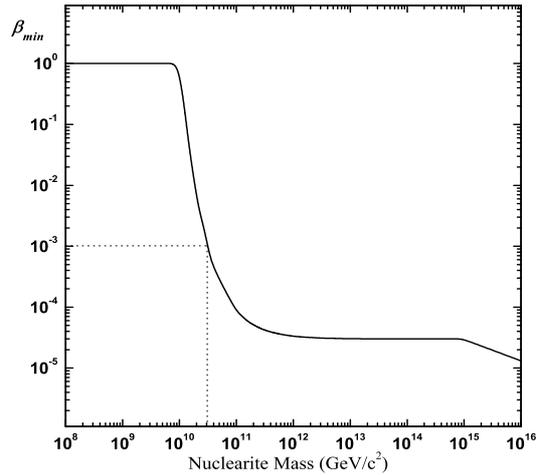}}
\caption{The minimum velocity $\beta$ versus mass $M$ for nuclearites to be detected by the SLIM experiment at the Chacaltaya Level ($540 \textrm{g/cm}^{2}$).}
\label{fig:acceptance}
\end{figure}

Their energy loss is only dependent on the nuclearite radius and velocity and is given by (De Rujula and Glashow, 1984, De Rujula 1985):
\begin{equation}
\label{eq:eloss}
\frac{dE}{dx}=-\sigma \rho v^{2}
\end{equation}
where $\rho$ is the density of the traversed medium, $v$ is the nuclearite velocity and $\sigma$ is its cross section:
\begin{figure*}
\begin{center}
\resizebox{!}{6.8cm}{\includegraphics{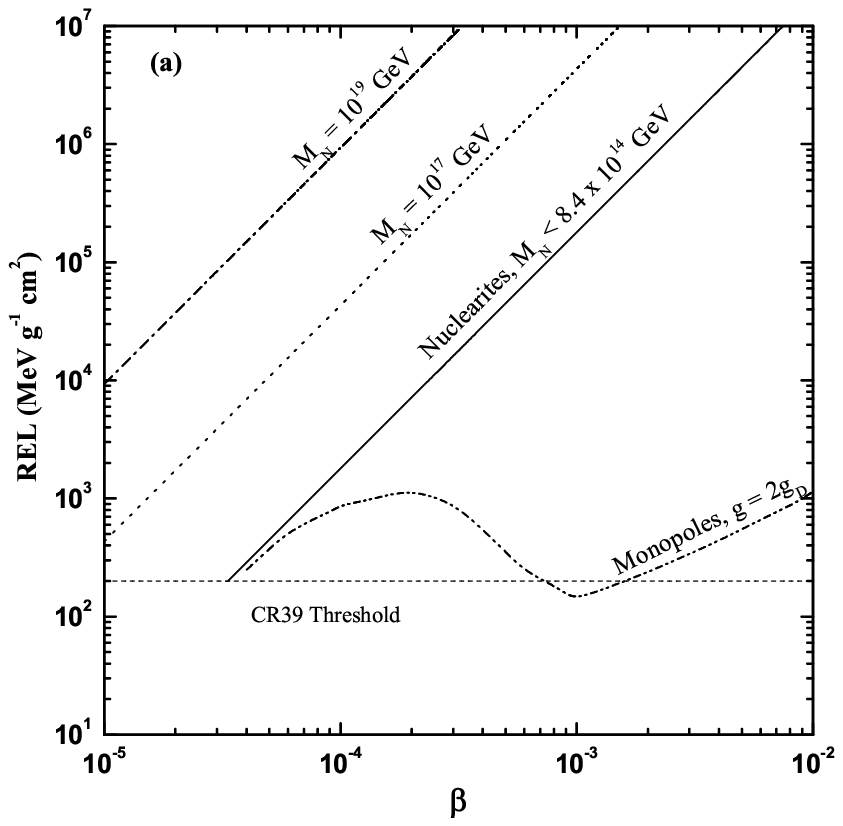}}
\resizebox{!}{6.85cm}{\includegraphics{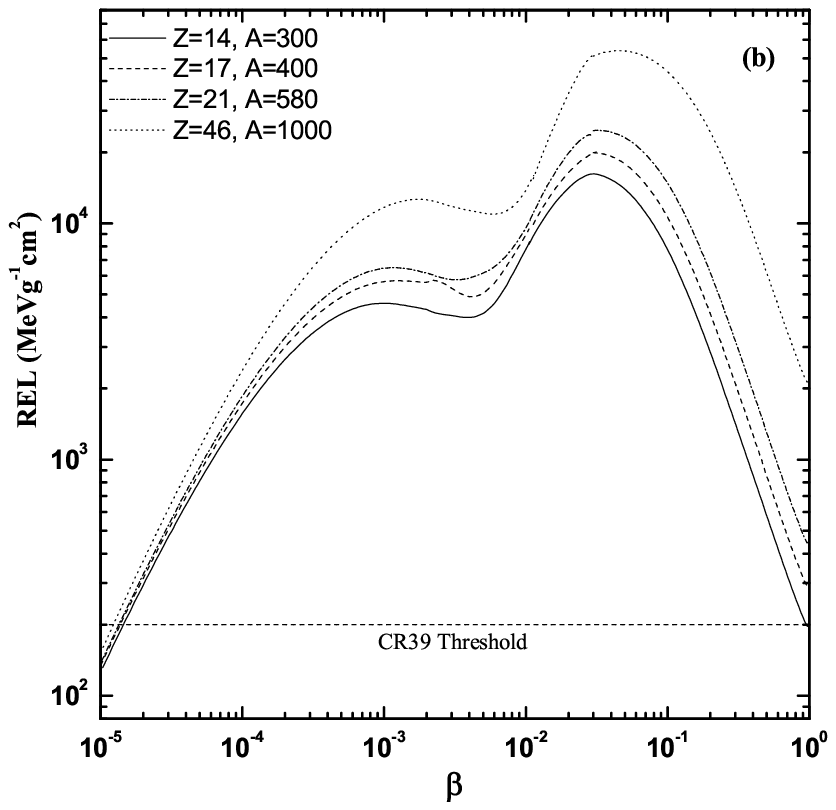}}
\caption{(a) REL versus $\beta$ for nuclearites of different masses. The REL of magnetic monopoles with charge $g$$=2g_{D}$ is also shown for comparison.
(b) REL versus $\beta$ for strangelets with different charges and masses. The strong etching threshold for CR39 is indicated by the horizontal line.}
\label{fig:REL}
\end{center}
\end{figure*}

\begin{equation}
\label{eq:sigma}
\sigma =\left\{ \begin{array}{ll}
\pi \cdot {10}^{-16} \,\textrm{cm$^{2}$}     & \qquad \text{for $ M < 1.5 \,$ ng}\\
 & \\
\pi {\left(\displaystyle{\frac{3 M}{4 \pi {\rho}_{N}} }\right)}^{2/3}     & \qquad \text{for $M > 1.5 \,$ ng}\
\end{array} \right.
\end{equation}
$\rho_{N}=3.5\,10^{14} \textrm{g/cm}^{3}$ is the nuclearite density and $M$ its mass.
Fig.~\ref{fig:REL}a shows the Restricted Energy Loss (REL) of nuclearites in CR39 nuclear track detector versus their initial velocity for different masses. The accessible region for the SLIM experiment at the Chacaltaya level ($540$ g/cm$^{2}$ of residual atmosphere) is expressed as the minimum incident velocity of nuclearites at the top of the atmosphere yielding a detectable signal at the detector level. For typical galactic velocities of $10^{-3}$ c, the SLIM experiment was able to detect any downgoing nuclearite with a mass larger than $3\cdot10^{10} \textrm{GeV/c}^{2}$, Fig.~\ref{fig:acceptance}.

Smaller SQM bags ($A$$<$$10^{7}$) are called Strangelets. They could be produced in the core of neutron stars and released into the Galaxy in strange star collisions (Friedman and Caldwell, 1991) and supernovae explosions (Vucetich and Horvath, 1998).
Strangelets are expected to undergo the same acceleration and interaction processes as ordinary cosmic rays. Their energy losses in matter should also be similar to those of heavy ions, Fig.~\ref{fig:REL}b; but they should have a much smaller charge to mass ratio compared to ordinary nuclei:
\begin{equation}
\label{eq:mass}
\begin{array}{ll}
Z/A\sim 0.1   & \qquad \text{for} \quad A\ll 10^{3} \\
  & \\
Z/A\sim 8~A^{-2/3}  & \qquad \text{for} \quad A\gg 10^{3}\
\end{array}
\end{equation}
as from Heiselberg, 1993.
\begin{equation}
\label{eq:cflmass}
Z/A\sim 0.3~A^{-1/3}
\end{equation}
as from Madsen, 1998.

Strangelets are also expected to have a high penetrability into the atmosphere; two different propagation models predicted their detection probability in high altitude experiments with a flux of the order of $10^{-13}$ cm$^{-2}$s$^{-1}$sr$^{-1}$ (Wilk et al., 1996, Rybczynski et al., 2001, 2006, Banerjee et al., 1999, 2000a, 2000b). The first propagation model assumes that relativistic strangelets decrease in mass when colliding with air nuclei until they reach a critical mass for neutron evaporation (Wilk et al., 1996). The second model, on the opposite, assumes that strangelets accrete neutrons and protons along their path increasing their mass and charge (Banerjee et al., 1999, 2000a, 2000b).

Other particles discussed are Q-balls. They are non topological solitons predicted by minimal supersymmetric extensions of the Standard Model of particle physics (Coleman, 1985, Kusenko et al., 1998a) and may be components of the cold dark matter. Q-balls are classified into two groups: supersymmetric electrically neutral solitons (SENS) and supersymmetric electrically charged solitons (SECS). While neutral Q-balls (SENS) may not be relevant to searches with nuclear track detectors because they mainly interact with matter by transforming nuclei into pions (Kusenko et al., 1998b, 2005), charged Q-balls (SECS) would interact with matter in a way not so different from SQM and thus could also be recorded by the SLIM experiment.
 
In the following we briefly present the apparatus and the experimental procedure. Then, we give the resulting flux upper limits for SQM nuggets and Q-balls set by SLIM.

\section{Experimental Procedure}
\begin{figure}
\centering
\resizebox{!}{0.2\textheight}{\includegraphics{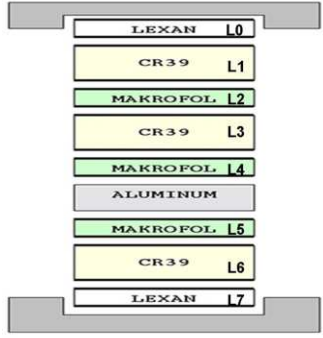}}
\caption{NTDs arrangement in a SLIM module.}
\label{fig:sketch}
\end{figure}

The SLIM experiment was an array of 7410 modules of Nuclear Track Detectors (NTDs) of $24\times24$ cm$^{2}$ area. Each module consisted of stacks composed of 3 layers of CR39$^{\scriptsize\textregistered}$ (Intercast Europe Co. Italy), 3 layers of Makrofol DE$^{\scriptsize\textregistered}$ (Bayer AG Germany), 2 layers of Lexan$^{\scriptsize\textregistered}$ and a 1mm thick aluminum absorber to slow down nuclear recoils, Fig.\ref{fig:sketch}.
SLIM NTDs were exposed to the cosmic radiation at the Chacaltaya high altitude Laboratory (5230 m a.s.l.) for 4.22 years; then they were brought back to the Bologna Laboratory where they were etched and analyzed.
When a charged particle crosses a nuclear track detector it produces damages along its trajectory forming the so-called latent-track. The following chemical etching leads to the formation of etch-pit cones in both front and back faces of the detector's foils provided the etching velocity $v_{T}$ along the track is larger than the etching velocity of the bulk material $v_{B}$. The response of the detector is given by the etching rate ratio $p$$=$$v_{T}/v_{B}$ which is dependent on the Restricted Energy Loss (REL) of the incident ion in the medium and thus on $(Z/\beta)^{2}$ with $Z$ and $\beta$ ($v$/c, c the velocity of light) the charge and velocity of the incident ion. The calibration with heavy relativistic ion beams (In$^{49+}$, Pb$^{82+}$ and Fe$^{26+}$) at accelerator facilities allows to build a data set of $(p$$-$$1)$ versus REL, Togo et al., 2008.
Nuclear tracks detectors were used in several applications for ex. the measurement of fragmentation cross sections (Giorgini et al., 2008) and also to search for highly ionizing radiation in underground laboratories (Ambrosio et al., 2002).
 Since the energy loss of the exotic particles of interest is constant in the detector modules the chemical etching of SLIM NTDs should have lead to the formation of collinear etch-pit cones with base areas equal within experimental uncertainties. Two etching processes were used to increase the CR39 signal to noise ratio; the strong etching: 8N KOH + 1.5\% ethyl alcohol at $75^{\circ}\mathrm{C}$ for 30 hours and the soft etching: 6N NaOH + 1\% ethyl alcohol at $70^{\circ}\mathrm{C}$ for 40 hours (Balestra et al., 2007, Manzoor et al., 2005, Cecchini et al., 2001b). The resulting thresholds were $Z/\beta$$\sim$$14$ ($REL_{th}$$\sim$$200$ MeVg$^{-1}$cm$^{2}$) in strong etching conditions and $Z/\beta$$\sim$$7$ ($REL_{th}$$\sim$$50$ MeVg$^{-1}$cm$^{2}$) in soft etching conditions. For Makrofol the etching conditions were 6N KOH + 20\% ethyl alcohol at $50^{\circ}\mathrm{C}$ for 10 hours and the threshold was at $Z/\beta$$\sim$$50$ ($REL_{th}$$\sim$$2.5$ GeVg$^{-1}$cm$^{2}$).
All uppermost CR39 layers (L1 in Fig.\ref{fig:sketch}) of SLIM's modules were etched in strong conditions, then scanned through binocular optical microscopes looking for a twofold coincidence between equally sized front and back cones. If a candidate was found the bottom CR39 layers (L6) were etched in soft conditions and scanned in a small square area around the expected position searching for a possible correspondence. This occurred in about 10\% of the cases. The third CR39 foils (L3) were etched only in few cases to check for possible uncertainties. A few Makrofol foils were also etched for similar reasons. More information on calibration procedure and analysis can be found in Balestra et al., 2008a, Cecchini et al., 2008, Medinaceli et al., 2008 and Togo et al., 2008.

No candidate event remained from all cross checks. Two very unusual events were observed but they were finally classified as manufacturing defects (Balestra et al., 2008b).

\section{Results and Discussions}
\begin{figure}[b!]
\centering
\resizebox{7.1cm}{6.7cm}{\includegraphics{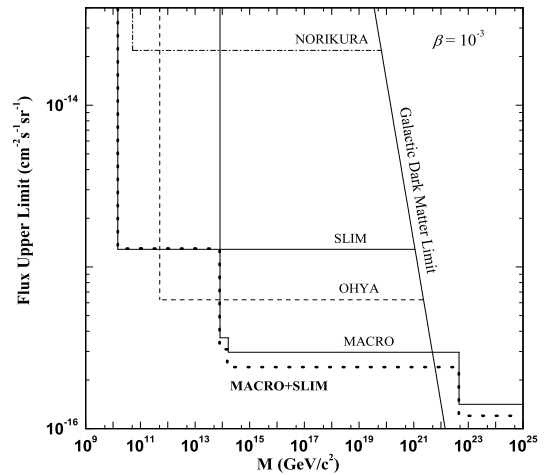}}
\caption{90\% C.L. flux upper limits versus mass for nuclearites with $\beta$$=$$10^{-3}$ set by the SLIM and other NTD experiments. The combined MACRO+SLIM upper limit is also shown.}
\label{fig:NuclLimitComp}
\end{figure}
The analysis of $427$ m$^{2}$ of Nuclear Track Detectors exposed to the cosmic radiation for 4.22 years at the Chacaltaya high altitude laboratory yielded no event. This allowed to set new upper limits for downgoing Strange Quark Matter nuggets and charged Q-balls, of the order of $1.3\cdot10^{-15}$ cm$^{-2}$s$^{-1}$sr$^{-1}$ at the 90\% C. L.

\begin{figure}
\centering
\resizebox{7.1cm}{6.7cm}{\includegraphics{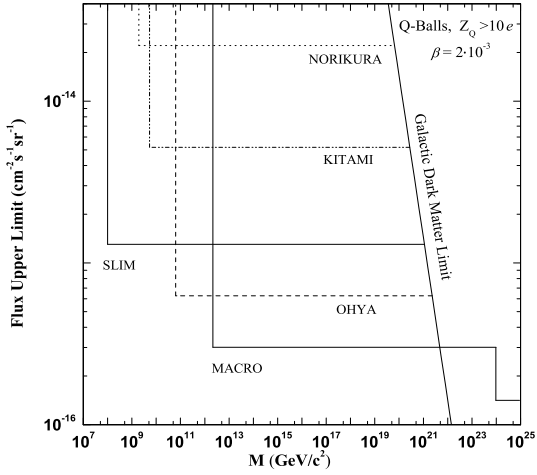}}
\caption{90\% C.L. flux upper limits versus $M$ for a downgoing flux of charged Q-balls with charges $Z_{Q}$$>$$10e$ and velocity $_beta$$=$$2\cdot10^{-3}$, set by various experiments using NTDs.}
\label{fig:QballLimitComp}
\end{figure}

 In Fig.~\ref{fig:NuclLimitComp} are shown the flux upper limits for nuclearites with velocity $\beta$$=$$10^{-3}$ set by different experiments using NTDs: MACRO (Ambrosio et al., 2002), OHYA (Orito et al., 1991) and NORIKURA (Nakamura et al., 1991.). These limits can be extended to charged Q-balls as shown in Fig.~\ref{fig:QballLimitComp} for charges $>$$10e$ and velocities of the order of $2\cdot10^{-3}$. Results of other experiments with NTDs are also shown for comparison (Ambrosio et al., 2002, Orito et al., 1991, Nakamura et al., 1991, Doke et al., 1983, KITAMI).\\
Note that in the case where charged Q-balls accrete electrons their energy losses are about the same as for nuclearites thus, the conclusions obtained for nuclearites can be extended to them.

\begin{figure}
\centering
\resizebox{7.5cm}{6.7cm}{\includegraphics{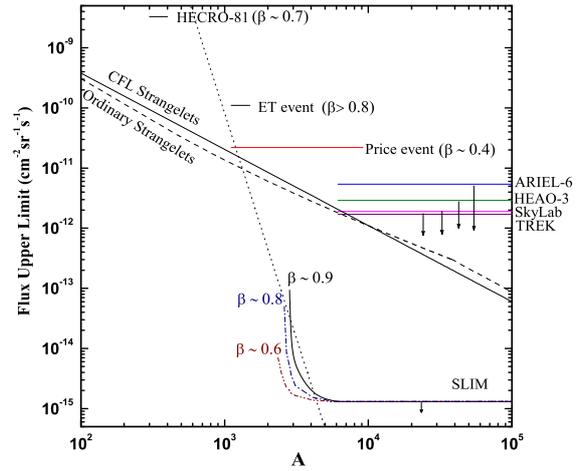}}
\caption{90\% C.L. flux upper limits versus baryon number A for light downgoing strangelets set by SLIM and experiments onboard balloons and satellites, see text. The lines are the expected fluxes based on different models: solid and dashed lines are for two types of strangelets produced in strange star collisions (Madsen, 2005) and dotted line as evaluated by Rybczynski et al. (2001, 2006).}
\label{fig:StrLimitComp}
\end{figure}

The same limit applies for downgoing strangelets with velocities at the detector level $\beta$$<$$0.6$, as can be seen from Fig.~\ref{fig:StrLimitComp}. Other limits derived from experiments onboard balloons and in space are also shown for comparison (Fowler et al., 1987, ARIEL-6; Binn et al., 1989, HEAO-3; Shirk and Price, 1978, SkyLab; Westphal et al., 1998, TREK).  Lines are flux predictions from different models (Madsen, 2005; Rybczynski et al., 2001, 2006). The fluxes of some cosmic ray unexpected events are also shown (Sandweiss, 2004). In order to be able to compare our results with other experiments we had to assume a particular propagation model through the Earth atmosphere.
According to the model of decreasing strangelet when propagating through the atmosphere (Wilk et al., 1996, Rybczynski et al., 2001, 2006) the SLIM experiment with its $427$ m$^{2}$ area and 4.22 years exposure should have detected about 5 to 10 events, taking into account the detector angular acceptance. In the case of the propagation model assuming strangelets growing in mass and charge (Banerjee et al., 1999, 2000a, 2000b) the number of events that should have been recorded by SLIM is about 80 events. The null observation could thus exclude this model while the model of decreasing strangelet can only be marginally excluded.
\section{Conclusions}
No candidate event was found after more than four years exposure of the SLIM experiment at the high altitude Chacalataya laboratory.  New upper limits were set for the fluxes of downgoing SQM nuggets and charged Q-balls at the level of $1.3 10^{-15}$ cm$^{-2}$s$^{-1}$sr$^{-1}$. The SLIM experiment has extended the search for intermediate mass SQMs and Q-balls and constrained models for strangelet propagation through the Earth atmosphere.

\section*{Acknowledgments}
We acknowledge the collaboration of S. Balestra, E. Bottazzi, L. Degli Esposti and G. Grandi of INFN Bologna and the technical staff of the Chacaltaya Laboratory. We thank INFN and ICTP for providing grants for non-italian citizens.

\end{document}